\documentclass[preprint]{aastex}

\usepackage{graphicx}
\usepackage{amsmath}
\usepackage{upgreek}
\usepackage{natbib}

\title{The {\sc xdspres} CL-based package for reducing OSIRIS cross-dispersed spectra}

\shorttitle{The {\sc xdspres} package}
\shortauthors{Ruschel-Dutra et al.}

\author{Daniel Ruschel-Dutra$^\dagger$, Rog\'erio Riffel, Jorge Ricardo Ducati and Miriani Pastoriza}
\affil{Departamento de Astronomia, Instituto de F\'isica, Universidade Federal do Rio Grande do Sul}
\email{$^\dagger$daniel.ruschel@ufrgs.br}

\accepted{July 6, 2011}

\date{July 6, 2011}

\cpright{ASP}{2011}

\slugcomment{Accepted for publication by PASP on 07/06/2011.}

\begin{document}

\begin{abstract}
We present a description of the CL-based package {\sc xdspres}, which aims at being a complete reducing facility for cross-dispersed spectra taken with the Ohio State Infrared Imager/Spectrometer, as installed at the SOAR telescope. This instrument provides spectra in the range between 1.2$\upmu$m and 2.35$\upmu$m in a single exposure, with resolving power of R $\sim$ 1200. {\sc xdspres} consists of two tasks, namely \textit{xdflat} and \textit{doosiris}. The former is a completely automated code for preparing normalized flat field images from raw flat field exposures. \textit{Doosiris} was designed to be a complete reduction pipeline, requiring a minimum of user interaction. General steps towards a fully reduced spectrum are explained, as well as the approach adopted by our code. The software is available to the community through the web site \textit{http://www.if.ufrgs.br/$\sim$ruschel/software}.
\end{abstract}

\maketitle

\section{Introduction}

Cross-dispersed spectroscopy makes possible to acquire information of wide spectral regions in a single exposure, by projecting several dispersion axes on the detector simultaneously. As a consequence, the reduction process required to analyze this kind of data is complicated, since different diffraction orders need to be selected, extracted, calibrated independently and combined in the final step. This difficulty led many authors to develop methods and software packages for the reduction of cross-dispersed and echelle spectra \citep[e.g.][]{moreno1982, rossi1985, piskunov2002, bochanski2009}.

In the past decade the near infrared (NIR) has also been explored by cross-dispersed spectrographs, such as Spex \citep{rayner2003} at the NASA Infrared Telescope Facility (IRTF), with a resolving power of $\sim$ 2000 and reaching from 0.8 to 5.5$\upmu$m. Other examples are TripleSpec \citep{edelstein2007} and the Folded-port Infrared Echellette (FIRE) \citep{simcoe2008}, achieving R $\sim$ 2600 and R $\sim$ 6000 respectively, and covering roughly the same wavelength domain (0.8 - 2.4$\upmu$m).

Another instrument of similar capabilities is the Ohio State Infrared Imager/Spectrometer (OSIRIS), currently installed at the Southern Astrophysics Research Observatory (SOAR), attached to the 4.1m telescope. OSIRIS provides spectral coverage from 1.0$\upmu$m to 2.4$\upmu$m in cross-dispersed mode, with a resolving power of $\sim$ 1200. High resolution (R $\sim$ 3000) long-slit modes are also available, but multi-band spectroscopy of this kind suffers from  differences in aperture and seeing.

However, reduction of NIR spectra has a complexity of its own, mostly related to telluric spectral features, both in absorption and emission, and black body radiation due to the telescope itself. A rich literature has been developed on the subject \citep[e.g.][]{maiolino1996,vacca2003,cushing2004}.

There are currently no specific software packages available for the reduction of cross-dispersed spectra taken with OSIRIS. Aiming at providing a fast and highly automated task, we developed the {\sc xdspres} (acronym for cross-dispersed spectra reduction script) package. The CL language was chosen due to the availability of almost all of the basic tasks needed to perform the reduction in the Image Reduction and Analysis Facility ({\sc IRAF}) software \citep{IRAF1,IRAF2}.

In \S \ref{sec:osiris} we describe main aspects of the instrument, focusing on its effects on the reduction process. In \S \ref{sec:reduction} we describe the general steps towards a fully reduced spectrum, as well as the approach adopted by the {\sc xdspres} package to each of these steps, and finally in \S \ref{sec:summary} we give a brief summary.

\section{OSIRIS}
\label{sec:osiris}

In this section we discuss the main aspects of the cross-dispersed mode of OSIRIS, with special attention to those characteristics that are relevant to the reduction process. A complete description of the instrument can be found in its on line User's Manual\footnote{http://www.ctio.noao.edu/instruments/ir\_instruments/ \\ osiris2soar/manual/}.

The detector is a 1024x1024 HAWAII array \citep{hodapp1996}, sensitive to wavelengths of up to 2.5$\upmu$m. Equation \ref{eq:linearity} models the non-linear behavior of the array, which only becomes critical above 28,000 counts. Usually the detector is read only at the end of the integration, but since it can be read non-destructively different sampling methods could be implemented.

A residual image is sometimes seen, specially when bright sources are observed in acquisition mode. This means that eventually some of the first spectra taken after the target acquisition images have to be discarded. Residuals have approximately 2\% of the intensity of the original source, and it should not be a problem to science exposures that have typical counts below one thousand.

{\small
\begin{align}
\frac{\mbox{ADU}^\prime}{\mbox{ADU}} &= 1.00108 - 1.015777 \times 10^{-6} \mbox{ADU} \notag \\
& + 1.548099 \times 10^{-10}\mbox{ADU}^2 \notag \\
& - 1.945376 \times 10^{-15}\mbox{ADU}^3
\label{eq:linearity}
\end{align}}

\begin{figure}[ht]
\plotone{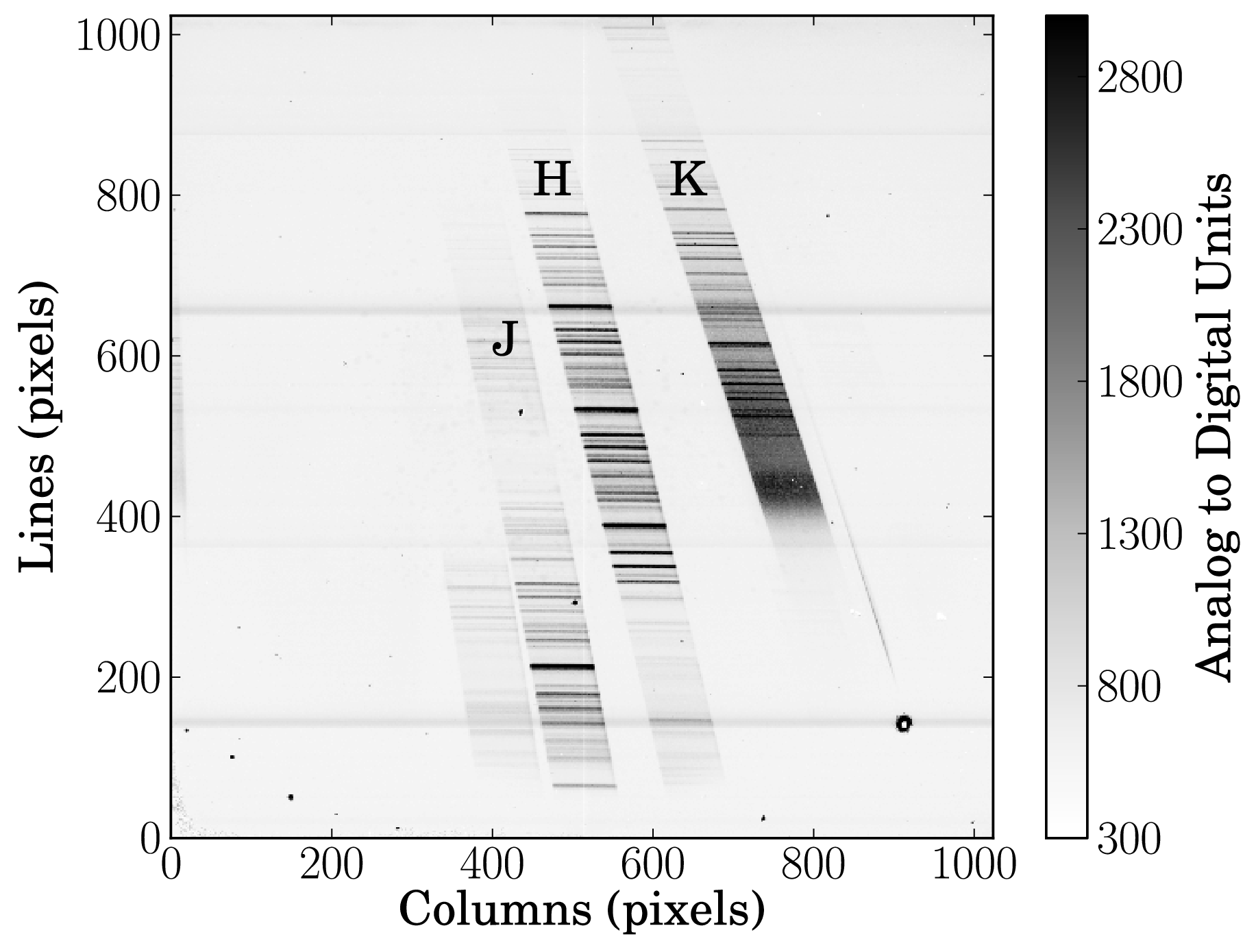}
\caption{Spectrum showing atmospheric emission lines and identifying the orders that are projected on the chip. The exposure time for this image was 20s. Horizontal lines seen in this and subsequent images, mainly around 140 and 650 pixels, are probably the result of scattered light that reached the grating.}
\label{fig:sky}
\end{figure}

In cross-dispersed mode OSIRIS projects almost six orders on the detector, from which three are extracted. Wavelength coverage for each of the extracted orders are 1.2 - 1.5, 1.5 - 1.9 and 1.9 - 2.35$\upmu$m, for the J, H and K bands respectively, all of them with R $\sim$ 1200. Orders that are not extracted include a small portion of the J band (1.0 - 1.2$\upmu$m), and second order duplicates of the J band, located to the right of the K band. Figure \ref{fig:sky} shows an example of sky spectrum where the three main orders are evident. Orders that are not extracted are also visible in figure \ref{fig:flats}.

From figure \ref{fig:sky} it can also be noted that dispersion axes are nearly vertical, meaning that within a given aperture each line corresponds to a particular wavelength. The misalignment between detector lines and wavelength coordinate are less than one pixel from one end of the slit to the other, or less than one third of the full width at half maximum (FWHM) of a emission line in the J band. Therefore corrections to dispersion axis orientation were not attempted, and all extractions assume a vertical dispersion.

\section{Reduction process}
\label{sec:reduction}

\subsection{Flat field}
\label{sec:flat}

\begin{figure}[ht]
\centering
\plottwo{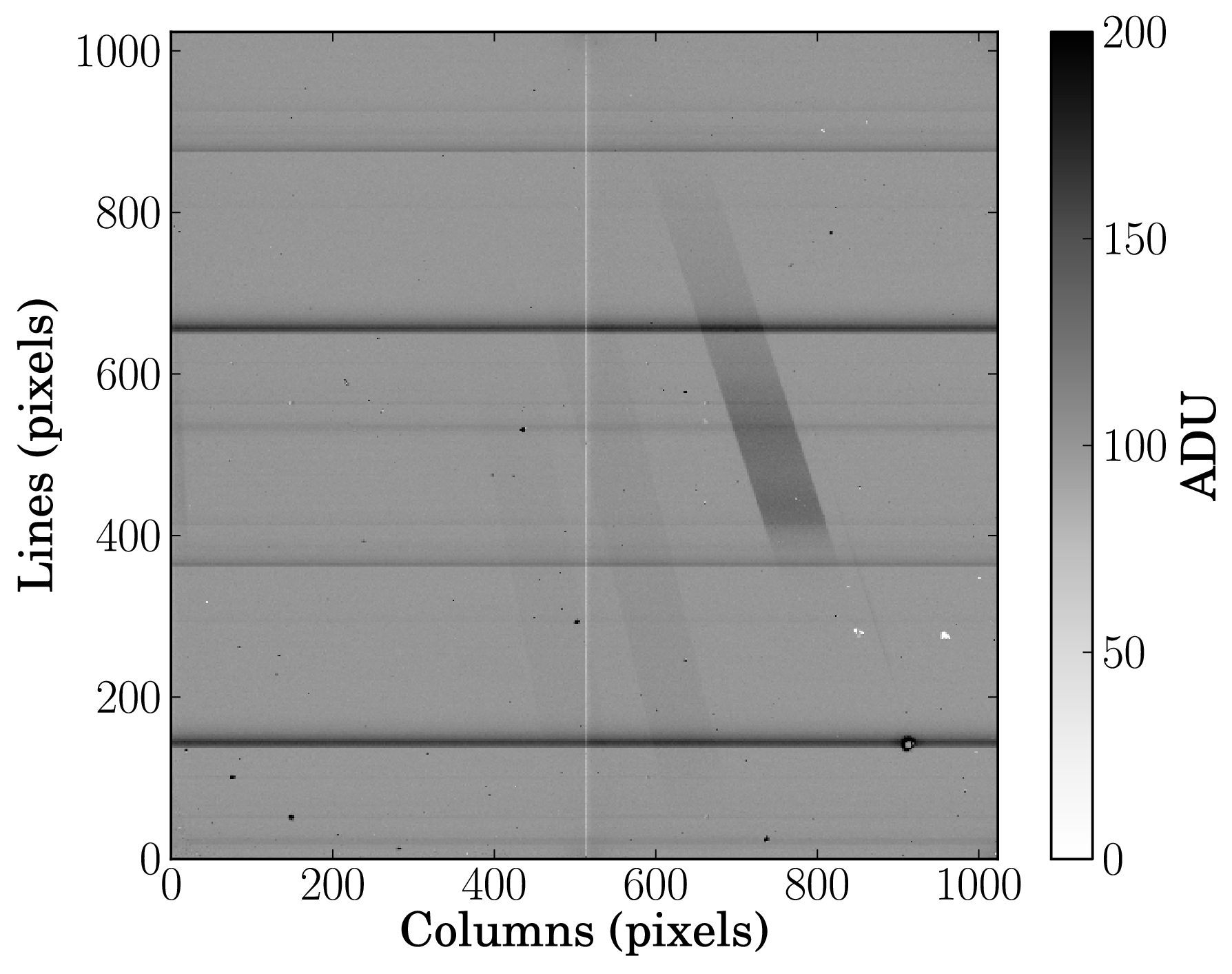}{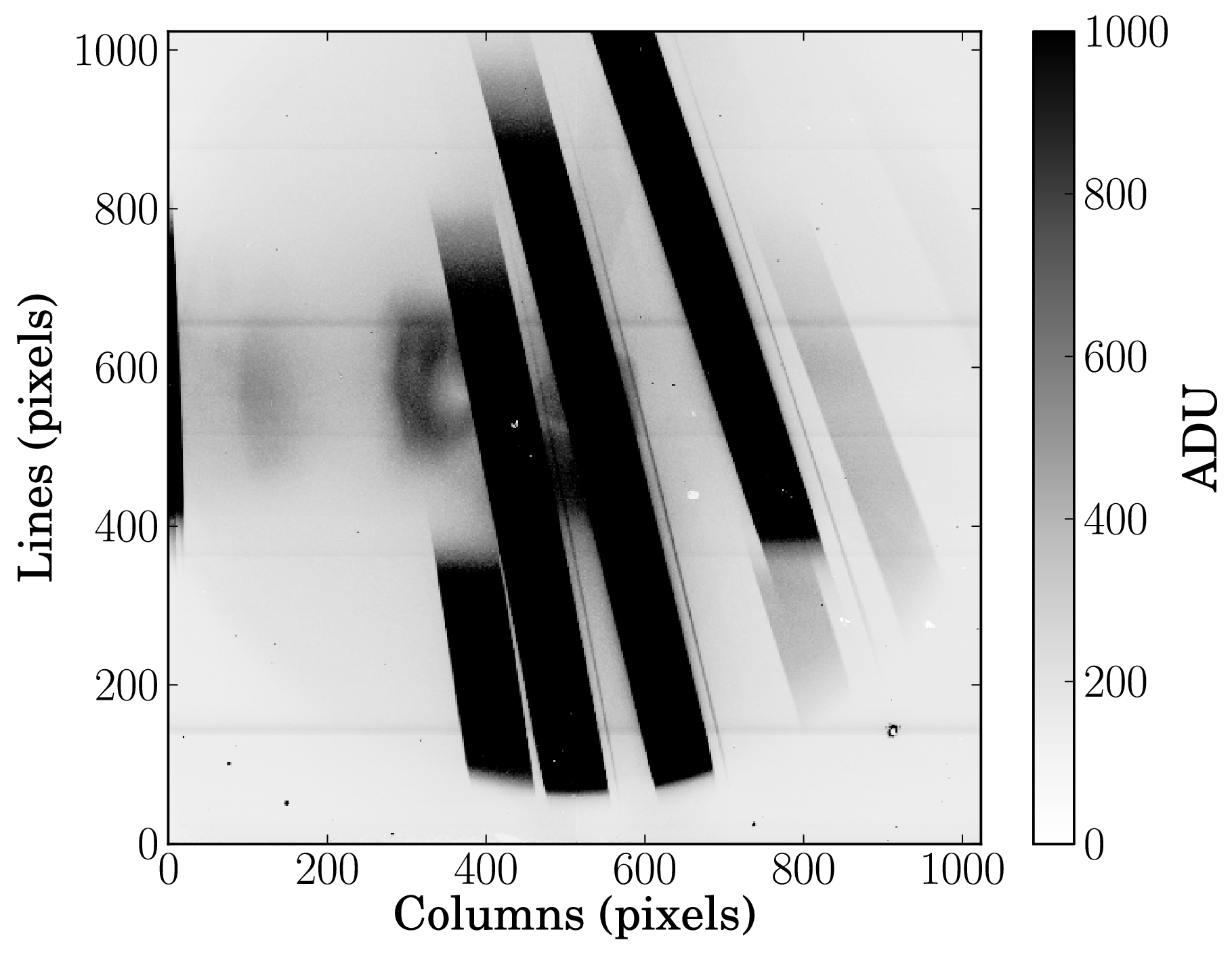}
\caption{\textit{Left} A sample flat field image with the lamp turned off. Thermal radiation from the telescope can be seen, as the flat field exposure is taken with grism already positioned. \textit{Right} A flat field with the lamp turned on, clearly showing the three orders in the center. Also visible are: further J band orders at the lower left and beyond the K band to the right; hot pixels in the lower corners; two groups of cold pixels near the center of chip; a small portion of an order at the detector's left border, between lines 400 and 800. Both images were taken with 3.2s of exposure.}
\label{fig:flats}
\end{figure}

In cross-dispersed mode, flat field images are taken with the cross-dispersing grism already positioned, which results in a spectrum of the flat field lamp, rather than an evenly illuminated image. Since the main purpose of a flat field is to identify pixel-to-pixel variations which are intrinsic to the detector, the continuum that corresponds to the spectral energy distribution of the lamp has to be removed. Moreover, two sets of flat fields are needed, one with the flat field lamp on and another with the lamp off. The later is required  because thermal radiation from the telescope becomes appreciable in the low energy end of the spectrum, as it can be seen in Fig. \ref{fig:flats}. Typical sets consist of 10 exposures of each kind.

The \textit{xdflat} task automates the preparation of a normalized flat field image, which will be later used to correct the science images. First it applies a linearity correction to all flat field images, according to equation \ref{eq:linearity}. Then both sets (flat-on and flat-off) are averaged independently, and the resulting flat-off image is subtracted from the flat-on. We have omitted a figure showing the subtracted flat because it is visually identical to the flat-on. The only noticeable difference is the suppression of a few hot pixels at the lower portion of the image.

To remove the spectrum of the flat field lamp \textit{xdflat} begins by extracting each order. Apertures are identified by a centering algorithm ({\sc apfind})  that searches for three local maxima in the central lines of the chip. The peaks are assumed to be separated by more than 30 pixels and have an approximate width of 80 pixels. Aperture sizes are reevaluated by setting the borders at 20\% of the peak intensity of each order. A tracing algorithm ({\sc aptrace)} moves in regular five pixel steps along the dispersion axis, assessing changes in peak location for each order, leading to a two-dimensional description of the aperture position. The aperture tracing function, a second order Legendre polynomial, is fitted to predefined sample regions of the chip that are less affected by scattered light. Errors in the two-dimensional aperture border definitions are usually below three pixels.

\begin{figure}[ht]
\centering
\plotone{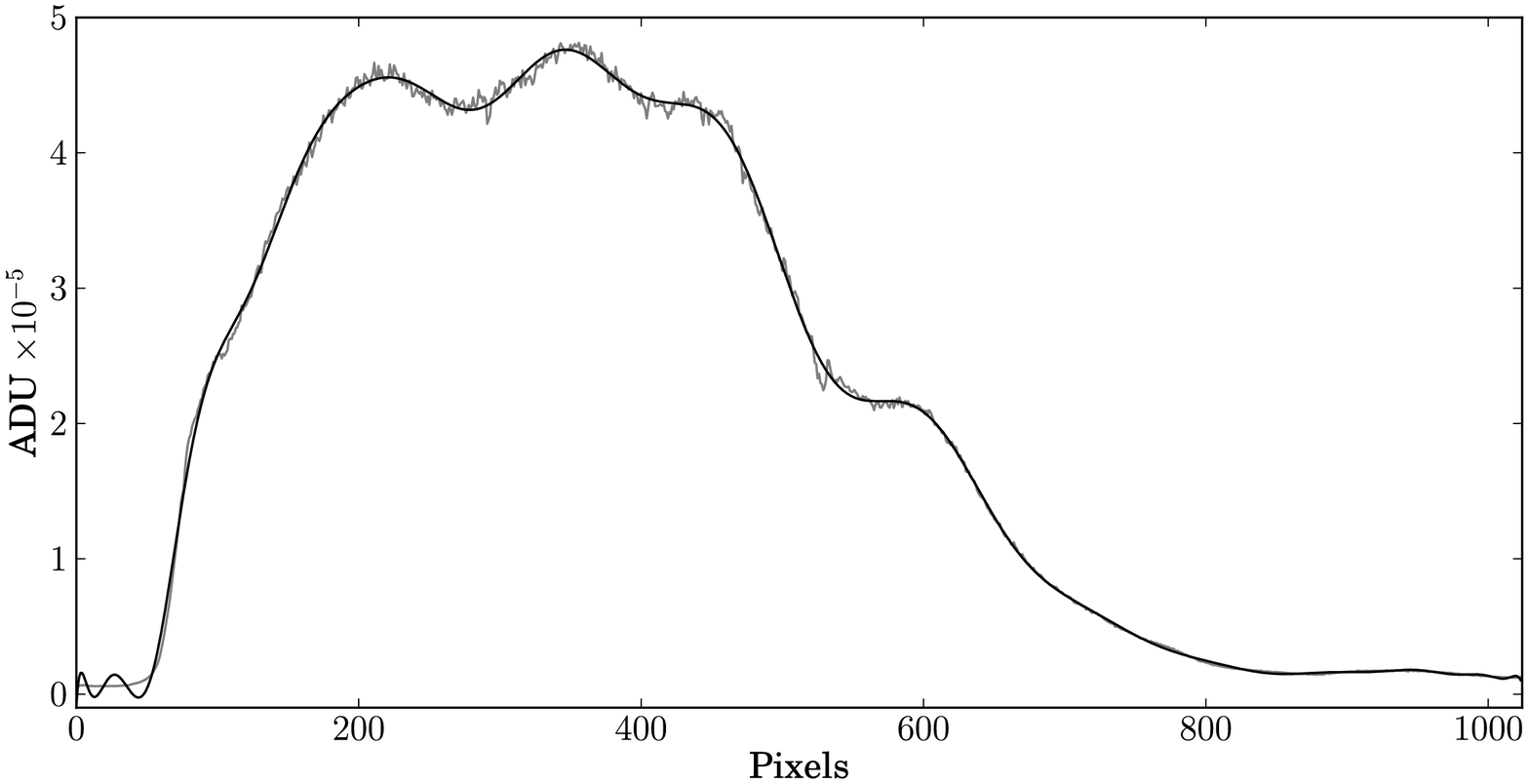}
\caption{Mean flat field spectrum at the J band (\textit{grey}), and fitted function (\textit{black}), RMS for this fit is $\sim$ 4600 ADU, which correspond to roughly 2\% of the average signal.  Artificial oscillations of the fit have no practical effects over the science spectrum as these portions are ignored after extraction.}
\label{fig:flatj}
\end{figure}

A 30th order Legendre polynomial is fit to the spectrum, which is then normalized. Such a high order polynomial is justified by the complex pattern produced by the flat field lamp as it passes through the spectrometer, as shown by figure \ref{fig:flatj}. Artificial oscillations at the apertures' limits are ignored after extraction. Typical RMS of the fit is below 5000 ADU, which may seem high but actually amounts to roughly 2\% of the average signal. The final flat-field image has all its pixel counts set to 1, except those on the regions occupied by the spectrum, which are replaced by the ratio between the original count and the fitted polynomial.

\subsection{Subtraction Object - Sky}

In the NIR spectral region the atmosphere plays an important role. Besides a significant telluric absorption, several atmospheric emission lines are entangled with the spectrum of the astronomical source (see figure \ref{fig:sky} for a sample spectrum of the sky, where the J, H and K bands are identified).

The process of removing telluric emission lines is commonly known as sky subtraction, or sky chopping, and the angular size of the target dictates whether additional off-source exposures are required. In the case of point sources, which occupy only a small fraction of the slit, one can take exposures with the source in two different positions along the slit, and later subtract subsequent images. This is the technique employed to obtain the spectra of standard stars, and it makes more efficient use of telescope time. When extended sources are concerned, a separate set of exposures taken from a nearby dark region of the sky is needed, a process commonly referred to as nodding.

The \textit{doosiris} task was developed to reduce spectra from extended sources, therefore it assumes that a set of sky exposures was taken along with the science exposures, in order to remove the telluric emission lines. There are two ways by which users can inform the software about the nature of each image, namely: interactively identifying them via \textit{SAO Image DS9}, or providing an ASCII file with the type of exposure with respect to its numerical order. For further details refer to the {\sc xdspres} Manual. No attempt was made to provide a software solution for identifying different types of exposures, as specific criteria regarding the spectrum of the astronomical target would have to be predefined, adding, in our judgement, unnecessary complexity to the code.

Nodding patterns that make best use of telescope time use each sky exposure in more than one subtraction, as in O-S-O or O-S-O-O-S\footnote{Where ``O'' stands for object and ``S'' for sky}. It is thus impractical to simply subtract a combination of sky images from an equivalent combination of target ones. Instead of assessing the relevant physical quantities, a routine searches for the best telluric calibrator based on the file name index, assuming that these are sequentially numbered after the time of exposure.

\subsection{Extraction and Wavelength Calibration}

Extraction of science spectra follows the same procedures that were described in section \ref{sec:flat}\footnote{Although \textit{doosiris} is prepared to automatically define the borders of each aperture, targets that have complex spatial profiles should be personally reviewed.}. The sky spectrum is extracted using the same aperture definitions of the target spectrum.

\begin{figure}[ht]
\plotone{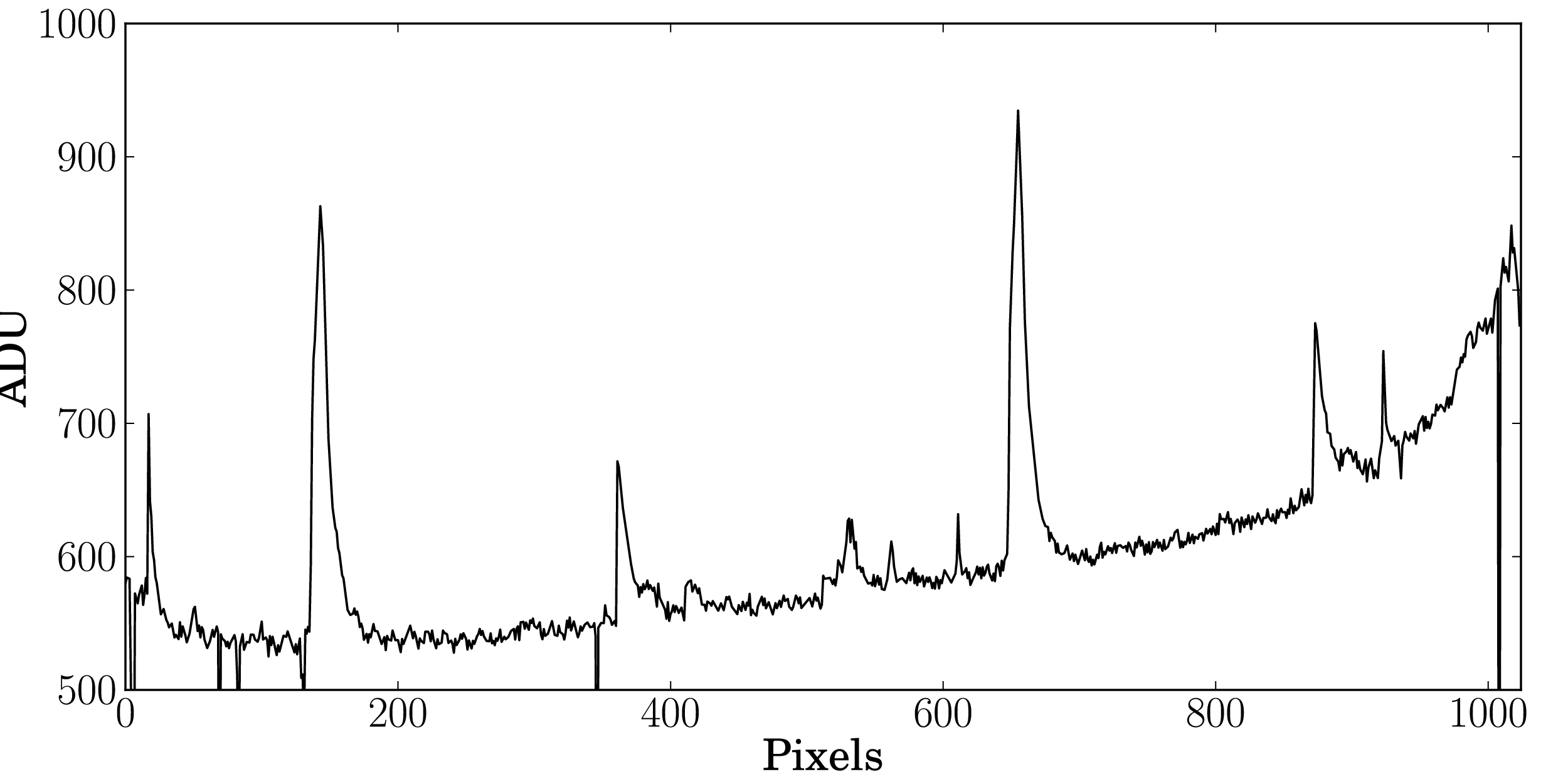}
\caption{Average of chip columns 980 to 1024. This background profile is removed prior to wavelength calibration to avoid confusion with OH emission lines.}
\label{fig:background}
\end{figure}

\begin{figure}[ht]
\centering
\plotone{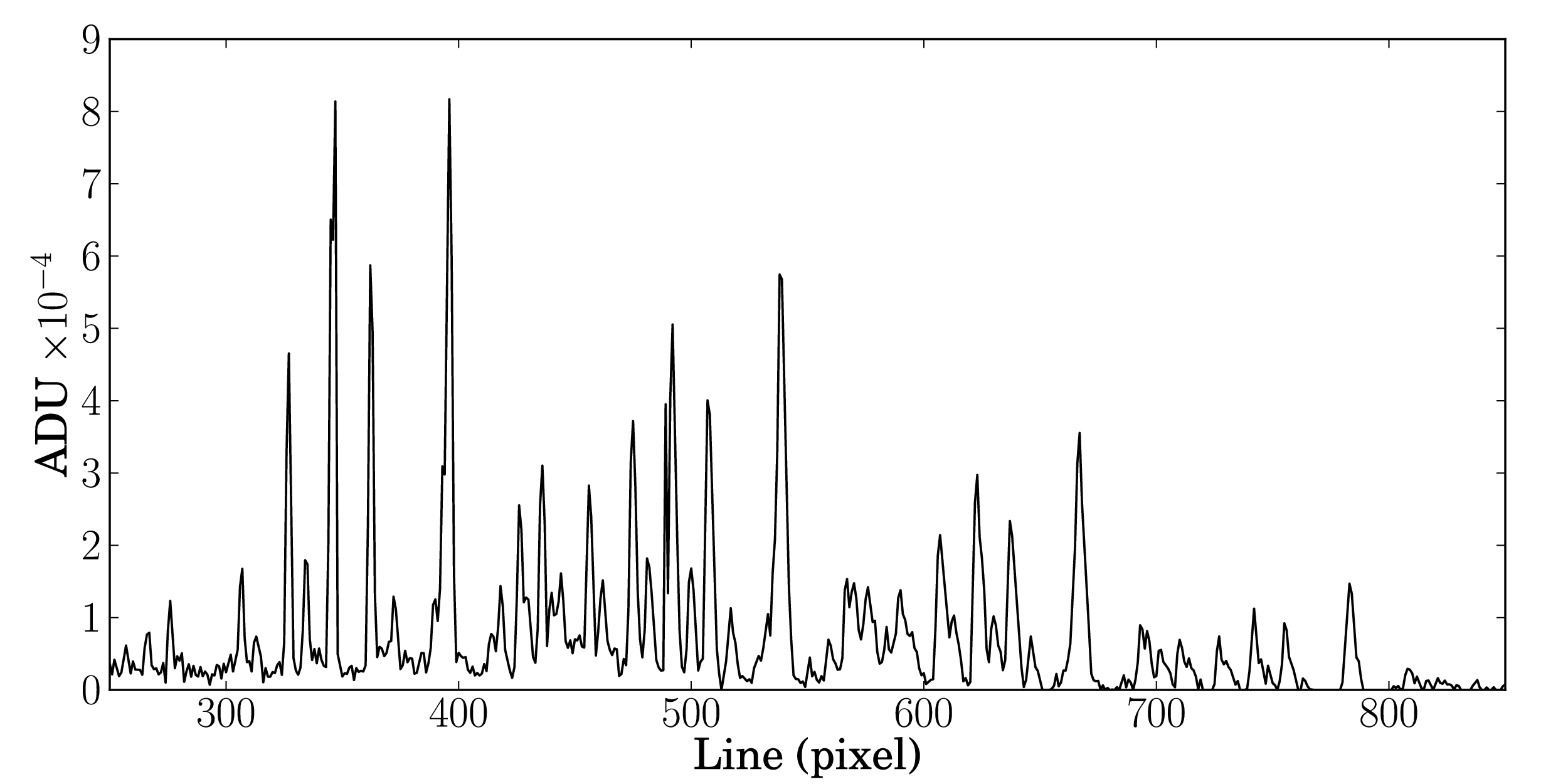}
\caption{A sample of the sky's spectrum at the H band.}
\label{fig:skyspec}
\end{figure}

Wavelength calibration is based on strong OH lines present in the sky exposures, a sample of which is shown in figure \ref{fig:skyspec}. As of the moment of the publication of this paper, OSIRIS presents what appears to be an illumination problem that produces lines across the detector, in the direction perpendicular to the dispersion axis. Since these lines can lead to confusion in the OH line identification process, a high order polynomial is used to fit and remove the vertical profile identified between columns 980 and 1024 (see figure \ref{fig:background}).

Interactive line identification is usually the best option, and since the dispersion function is almost linear there is no need for manually identifying more than four well spaced features. If the dispersion function fitting was successful, the unidentified features will match those in the line list provided with {\sc xdspres}, which was extracted from \citet{oliva1992}. \textit{Doosiris} also provides an option to automatically identify OH features in the spectrum of the sky that uses the \textit{reidentify} task, which requires a previously identified image. For $\sim$ 20 identified features, typical residuals are below 2 pixels, which translates into roughly $\pm$ 50 km s$^{-1}$.

\subsection{Telluric Removal and Flux calibration}
\label{sec:tell}

The subtraction of sky exposures from the on source images, obviously can only account for emission features of the atmosphere. To deal with the more subtle problem of removing atmospheric absorption \textit{doosiris} by default uses the spectrum of an A0V star, that should be obtained just before or after the science images. If the observed standard star has a different spectral type, the model atmosphere spectra, mentioned below, have to be replaced accordingly.

\begin{figure}[ht]
\centering
\plotone{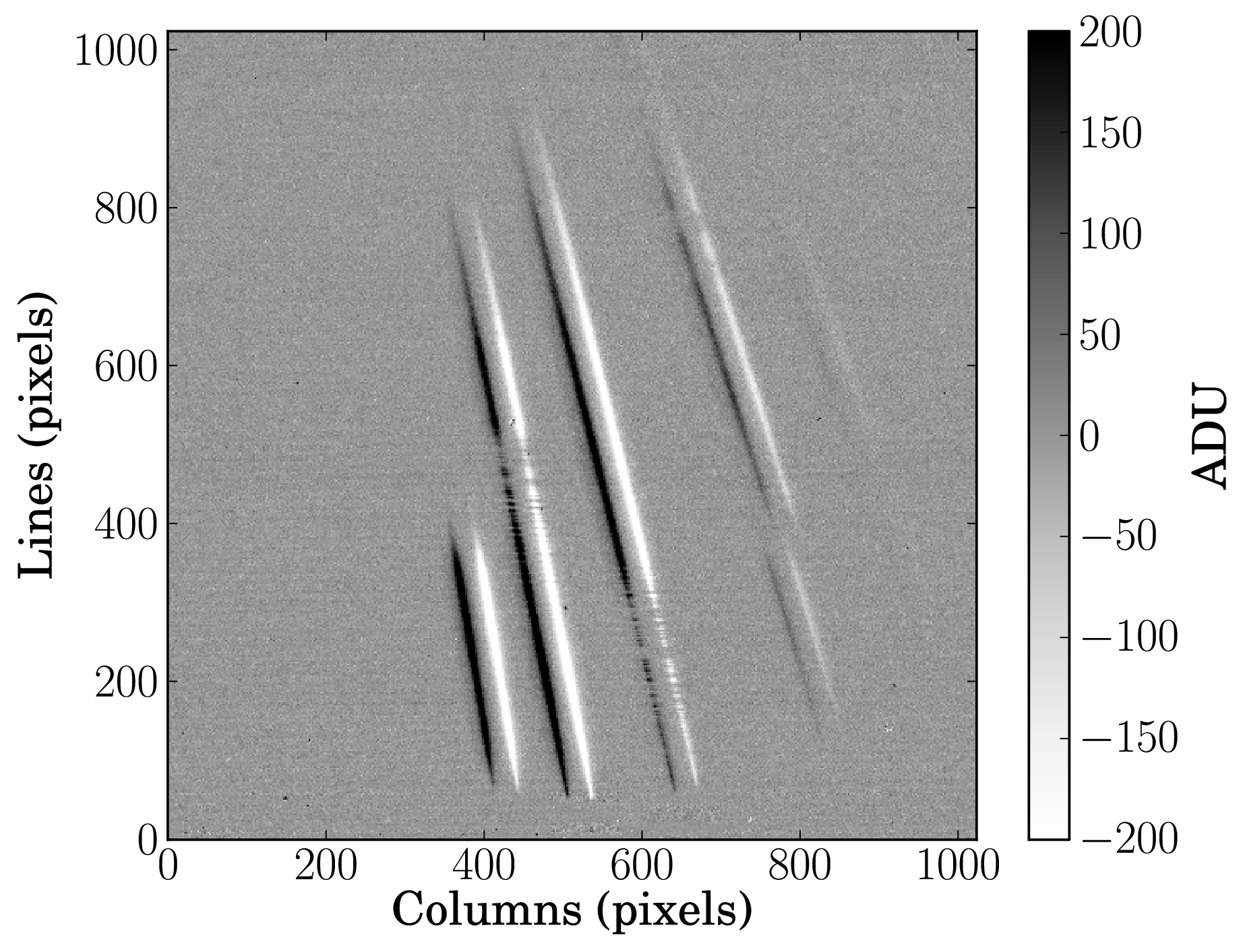}
\caption{The resultant image of subtracting subsequent exposures of the standard star occupying two different positions along the slit.}
\label{fig:star_sub}
\end{figure}

The standard star, being a point source, do not need a separate set of sky exposures, because it occupies only a small fraction of the slit. \textit{Doosiris} is prepared to manage two or three different star positions on the slit. In either case subsequent exposures are subtracted and the resulting images are summed; a sample of this sum can be seen in figure \ref{fig:star_sub}. After division by the normalized flat field image, both spectra are extracted and summed.

The spectrum of an A0V is almost devoid of metallic absorption lines, but the H lines that are present need to be eliminated before it can be applied to the science spectrum as a telluric calibrator. The method employed here follows the reasoning of \citet{vacca2003}, but with a different implementation. It basically consists in dividing the spectrum of the standard star by a model atmosphere of Vega, obtained from R. Kurucz\footnote{http://kurucz.harvard.edu/stars.html}.

\begin{figure}[ht]
\plotone{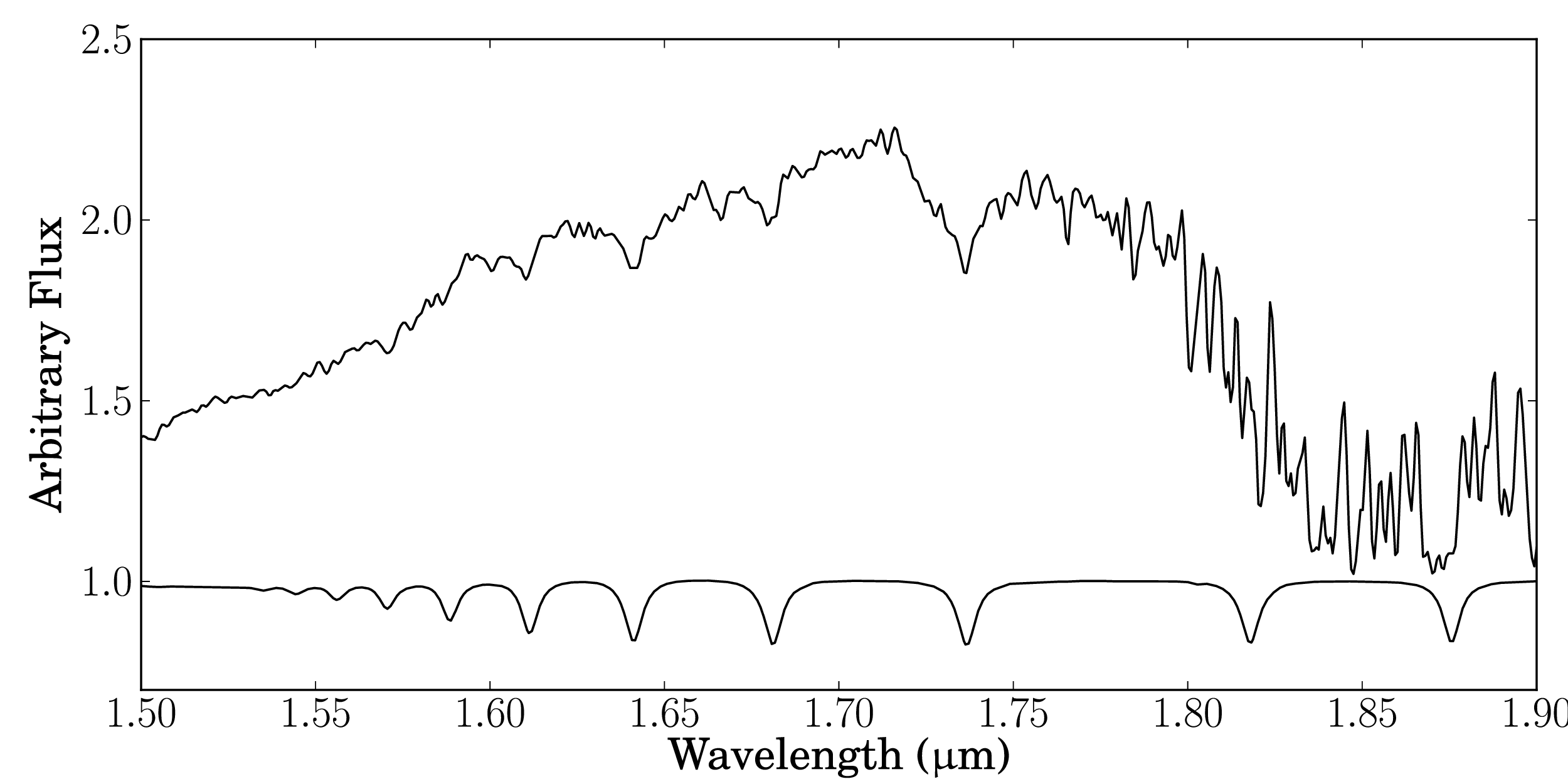}
\caption{Comparison between the standard star H band spectrum (top) and a model atmosphere (bottom) for Vega. The resolving power of both spectra is 1200.}
\label{fig:tell_star}
\end{figure}

First the model of Vega was smoothed by a Gaussian with $\sigma$ equal to the FWHM measured in a NeAr calibration lamp, to match the resolving power of the standard star. A spline was then adjusted to the continuum, leading to a purely absorption spectrum. The later is provided with the {\sc xdspres} package. The actual division of the reference star is performed by the \textit{telluric} task, which allows for the shifting and scaling of the model.  Figure \ref{fig:tell_star} shows a comparison between the observed spectrum of the standard star and a model atmosphere for Vega.

Once the absorption lines due to the stellar atmosphere have been removed, the spectrum becomes essentially a black body with telluric features. Its normalization by a polynomial that acts as a pseudo-continuum returns a purely telluric spectrum. Unabsorbed regions, that translate into sample regions for continuum fitting, were identified with the aid of NSO/Kitt Peak FTS data produced by NSF/NOAO\footnote{Available at http://www.eso.org/sci/facilities/paranal/ \\ instruments/isaac/tools/spectroscopic\_standards.html}. This division of the science spectrum also allows shifting and scaling. Some of the strongest telluric bands cannot be fully removed. Additionally, the high absorption in these regions causes a significant decrease in S/N.

\begin{figure}[ht]
\plotone{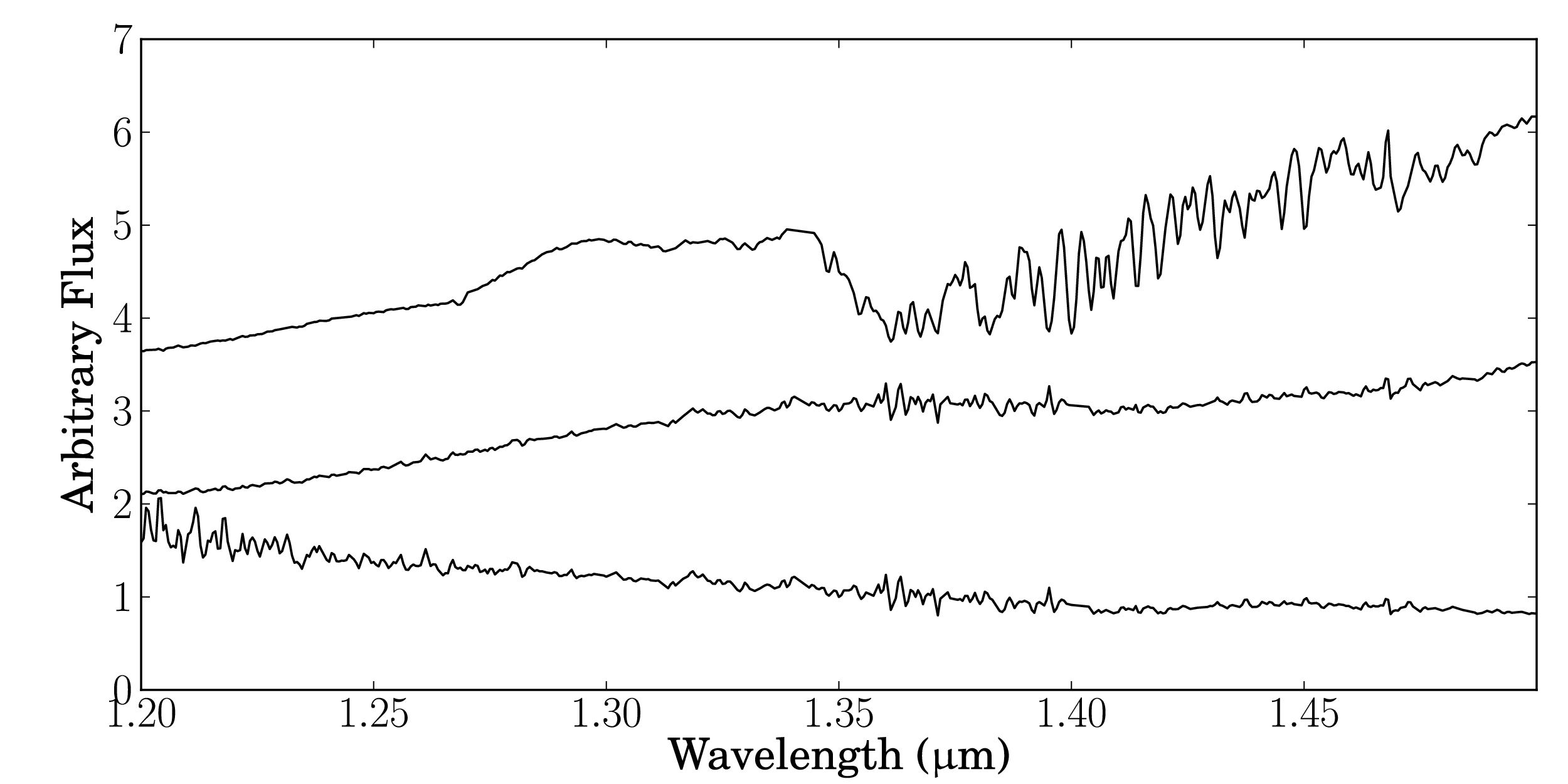}
\caption{From top to bottom: a) Sample spectrum at J band just after wavelength calibration. b) Same spectrum after the removal of telluric lines. c) Flux calibrated spectrum. Areas of strong telluric absorption cannot be fully corrected, and even if they could the signal to noise ratio would still be much lower than the rest of the spectrum.}
\label{fig:stages}
\end{figure}

The same polynomial employed as a pseudo-continuum for the reference star is later used to produce an independent sensitivity function for each aperture, by comparing it to a black body of 9480 K. This procedure restores the correct slope of the spectrum regardless of the accuracy in absolute flux. The later is estimated from the exposure time and magnitude of the standard star, which has to be provided by the user. Figure \ref{fig:stages} shows the effects of telluric line removal and flux calibration to a sample spectrum.

\begin{figure}[ht]
\plotone{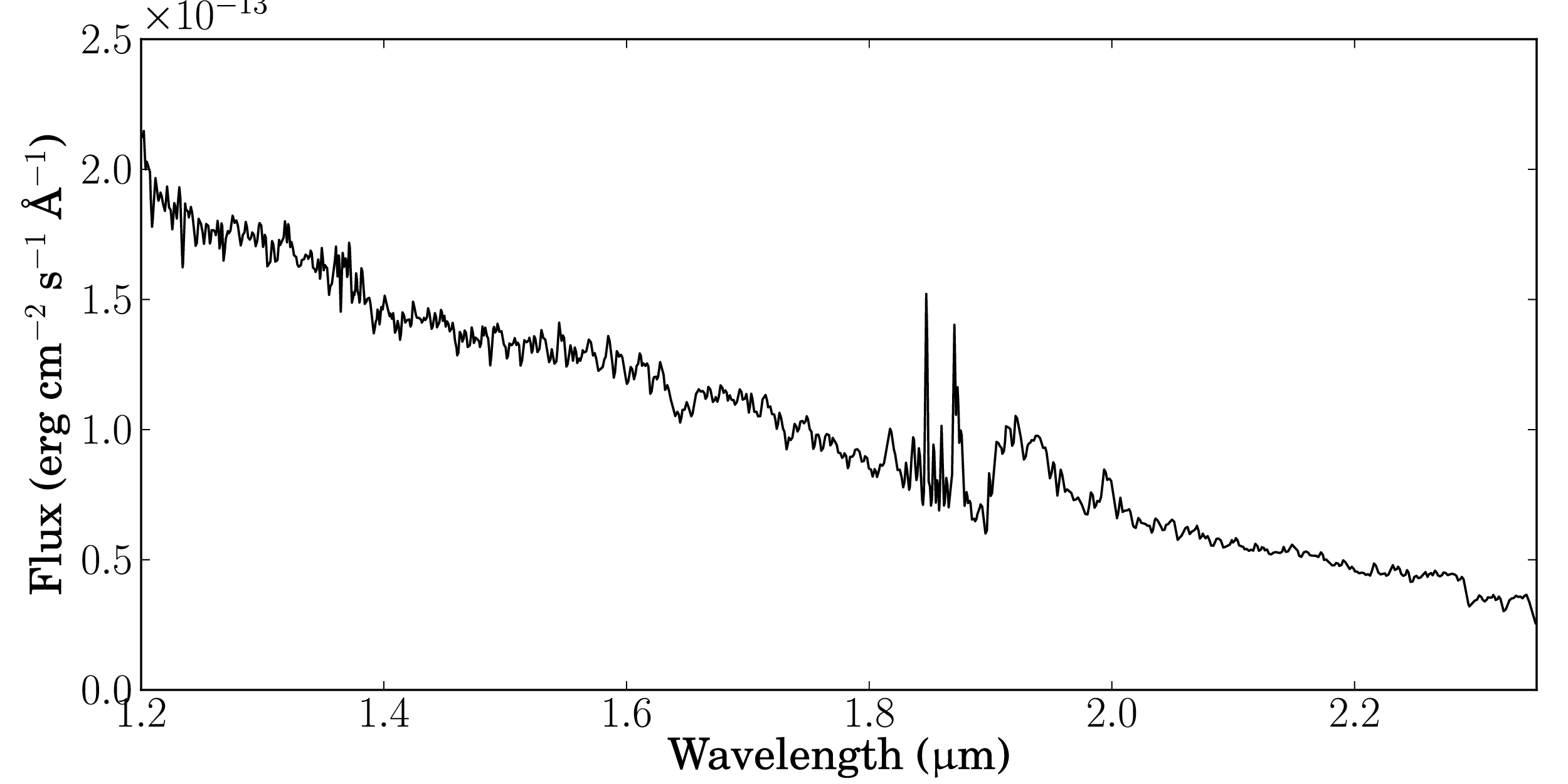}
\caption{Flux calibrated spectrum in the whole spectral range. Aperture transitions are at 1.5$\upmu$m and 1.9$\upmu$m. Strong telluric absorption bands that dominate the spectrum between 1.8 and 2.0$\upmu$m difficult the alignment between the H and K bands.}
\label{fig:complete}
\end{figure}

One would expect that a good flux calibration leads to a perfect alignment of the spectrum between different apertures. Although generally true, it has been observed that agreement is harder to achieve where the H and K bands meet. Strong telluric absorption bands near 1.9$\upmu$m difficult the evaluation of the sensibility function causing large deviations in the final spectrum. Figure \ref{fig:complete} shows a completely reduced spectrum encompassing the whole spectral range.

\section{Summary}
\label{sec:summary}

We have presented the {\sc xdspres} CL-based package, consisting of the \textit{xdflat} and \textit{doosiris} tasks, aimed at being a complete reduction facility for cross-dispersed spectra taken with the OSIRIS spectrometer, currently installed at the SOAR telescope. This particular instrument provides a relatively large spectral coverage, being able to project the full range between 1.2$\upmu$m and 2.35$\upmu$m over the detector in a single exposure. The blazing of different orders in the same image adds complexity to the already lengthy reduction of infrared spectroscopy data. {\sc xdspres} automatically performs the more mechanical and time consuming steps of the reduction, at the same time that it allows considerable user interaction in the more subjective stages. In addition, the possibility of a fast reduction provides means to make site adjustments to the observation strategy. As a sample of actually published data that was fully reduced with the {\sc xdspres} tasks, see \citet{riffel2011}. The complete software package and its documentation is available to the community at the web site \textit{http://www.if.ufrgs.br/$\sim$ruschel/software}.

\subsection*{Acknowledgments}

We thank an anonymous referee for very interesting comments that increased considerably the quality of the present paper. DRD thanks the support from the Brazilian research funding agency CNPq. OSIRIS is a collaborative project between the Ohio State University and Cerro Tololo Inter-American Observatory (CTIO) and was developed through NSF grants AST 90-16112 and AST 92-18449. CTIO is part of the National Optical Astronomy Observatory (NOAO), based in La Serena, Chile. NOAO is operated by the Association of Universities for Research in Astronomy (AURA), Inc. under cooperative agreement with the National Science Foundation. This work has been done with observations from the SOAR telescope, a collaboration among the Minist\'erio da Ci\^encia e Tecnologia/Brazil, NOAO, UNC and MSU.

\end{document}